Evolution of the "long tail" concept for scientific data

Gretchen R. Stahlman

School of Information, College of Communication & Information, Florida State University

Inna Kouper

Department of Informatics, Luddy School of Informatics, Computing, and Engineering, Indiana University - Bloomington

**Author Note**

Gretchen R. Stahlman ORCID ID: https://orcid.org/0000-0001-8814-863X

Inna Kouper ORCID ID: https://orcid.org/0000-0001-9801-2277

Correspondence concerning this article should be addressed to Gretchen Stahlman, 142 Collegiate Loop, Tallahassee, FL 32306-2100. Email: gstahlman@fsu.edu

**Abstract**

This review paper explores the evolution of discussions about "long tail" scientific data in the scholarly literature. The "long tail" concept, originally used to explain trends in digital consumer goods, was first applied to scientific data in 2007 to refer to a vast array of smaller, heterogeneous data collections that cumulatively represent a substantial portion of scientific knowledge. However, these datasets, often referred to as "long tail data," are frequently mismanaged or overlooked due to inadequate data management practices and institutional support. This paper examines the changing landscape of discussions about long tail data over time, situated within broader ecosystems of research data management and the natural interplay between "big" and "small" data. The review also bridges discussions on data curation in Library & Information Science (LIS) and domain-specific contexts, contributing to a more comprehensive understanding of the long tail concept's utility for effective data management outcomes. The review aims to provide a more comprehensive understanding of this concept, its terminological diversity in the literature, and its utility for guiding data management, overall informing current and future information science research and practice.

## Evolution of the "long tail" concept for scientific data

### Introduction

In statistics, the term "long tail" describes a distribution with a large number of low-probability events that have a small likelihood of occurring. Anderson (2007) applied this concept to digital consumer goods, observing that while a few items achieve high popularity, many more niche items have low demand but, cumulatively, they represent a substantial market share. The concept has been extended to scientific data, beginning with Palmer et al. (2007), who suggested that scientific data follows the "long tail" model where scientists generate many heterogeneous, smaller collections of data that represent significant institutional assets.

Heidorn (2008) expanded upon this work by pointing out the existence of "dark data" across the sciences, referring to data that remain underutilized due to improper indexing and storage, inadequate incentives, and lack of institutional and financial support for data management. He suggested that the distribution of research funding follows a power law, where a few well-funded projects (the "head") can afford robust data infrastructures and management, while many smaller, underfunded projects (the "tail") lack sufficient resources for data curation. Heidorn argued that if made accessible, long tail data would become a large voluminous source of new discoveries. The term "long tail data" has since been adopted by many scholars to refer to smaller data collections and to argue that as they comprise a substantial share of scientific data, they require significant attention along with improved policies, infrastructures, and training for their care (PLAN-E, 2018).

Nearly two decades have passed since the term "long tail" was first applied to describe scientific data. During this time, the socio-technical ecosystems surrounding data-oriented work have changed massively through development of new technologies for collaboration (Dai, Shin & Smith, 2018), emphasis on big data and artificial intelligence (Uzwyshyn, 2023), dissemination of open science initiatives (Gentemann, et al., 2022) and the widespread adoption of the FAIR principles for making data Findable, Accessible, Interoperable and Reusable (Wilkinson, et al., 2016). The 2023 Year of Open Science in the United States highlighted these shifts through a variety of programs to support sharing of research products including data (Gentemann, 2023). Despite all of these advances, barriers to *effective* data sharing continue to persist across disciplines (as a recent example, see Hughes, et al., 2023). In the continuing effort toward a better understanding and elimination of those barriers, synthesizing approaches to long tail data management and related dialogue is important because, as previously mentioned, researchers managing such data often must take substantial, often time-consuming curation actions themselves to make the data available to others.

This literature review therefore aims to inform relevant areas of Library & Information Science (LIS) research and practice about long tail data discourse and developments over time. As argued in Stahlman (2022), the intellectual intersection between data curation and human information behavior scholarship is fertile ground for harmonizing data centric and user centric approaches to better understand and model the lifecycles of long tail data. Along these lines, LIS researchers are beginning to articulate "data behavior" as a new research area and agenda (Zhang, et al., 2023), which is highly relevant for studying and facilitating long tail data management. Moreover, cultural shifts towards open science and data democratization bring significant information policy implications associated with long tail data. Organizational and institutional efforts to extract value from data while upholding FAIR principles and open science mandates highlight the intersection between data management, policy, and governance (Lane & Potok, 2024), which directly impacts handling of long tail data and potential tension points. In terms of data ethics, availability bias can occur where analysis is conducted using data that is readily available or abundant (Lund & Wang, 2022). It can also be argued that increased availability of long tail data can mitigate bias in data science by providing more robust representations of reality, which is an area for further study and consideration. Furthermore, LIS professionals and researchers are fundamentally concerned with developing services to satisfy the needs of users, and there remains a need for new services and frameworks that support researchers and enhance the discovery, access, and long-term viability of long tail data (Tenopir, et al., 2015).

This review paper overall contributes to the ongoing discussions of research data curation as it explores how the "long tail" concept and related framings of data have evolved since approximately 2007. Its purpose is not to explain the existence of long tail data; rather it aims to highlight the importance of integrating all research data in the efforts of open science and data curation and elevate the efforts of LIS practitioners and data managers tasked with supporting individual researchers and small teams and their data. Considering the changing and ever-expanding nature of data ecologies and their stewardship, an exploration of long tail data curation helps situate small data collections in the larger context of data paradigms and the entangled relationships between "big" and "small" data. A deeper understanding of this area of research and practice will also help bring together data curation discussions in LIS and in domain-specific contexts to inform current and future interdisciplinarity and collaborations. Finally, this review of the literature will contribute to a more holistic understanding of the long tail data discussions and synthesize their utility for successful data management outcomes in practice.

*A Note on Semantics*

The phrase "long tail" is a descriptor that has been applied to refer to smaller data collections that are numerous in aggregate across the sciences and require nuanced handling, along with the scientific practices that take place within the contexts of such data (PLAN-E, 2018). The dual meaning of the phrase is illustrated by Heidorn's (2008) conceptualization of a long tail funding distribution within which smaller data generated by individuals and project teams remain inaccessible in the lower-funded "tail" of the curve due to insufficient resources for curation. In this example, the "long tail" collections may be comprised of data such as biological specimens collected at a remote field station or data derived from larger archival datasets to produce publications, while the associated scientific practices may largely depend upon distribution and application of research funding. The literal meaning of the phrase similarly varies across disciplines and the broader literature, referring interchangeably to size of the data, visibility of data, impact of data, number of researchers working with datasets, and other characterizations (see Borgman, et al., 2016; Brooks, et al., 2016; Ferguson, et al., 2014; Liang, et al., 2010; Malik & Foster, 2012; Palmer, et al., 2007; Wallis, Rolando & Borgman, 2013). In other words, "long tail" is not simply a matter of scale and can also encompass larger datasets that receive little attention. For the purpose of this literature review, we synthesize these perspectives and consider the core definition of "long tail data" to be:

> *Vast amounts of distributed, heterogeneous, often smaller scientific datasets of various types generated by individual researchers or small research groups; these data do not receive the same level of attention, funding, or infrastructure support as larger, well managed datasets but nevertheless contain a wealth of valuable specialized information that can potentially contribute to scientific knowledge.*

The discussion around long tail data originated within the digital curation community, a growing subdiscipline of LIS, through which *data curation* has been defined as "the ongoing processing and maintenance of data throughout its lifecycle to ensure long term accessibility, sharing, and preservation" (NNLM, n.d.). Examples of long tail data include local ecosystem studies that inform environmental research when aggregated as well as niche surveys and small-scale studies in disciplines ranging from astronomy to social sciences. Given the dispersed use and multidimensionality of terminology surrounding long tail data, throughout this review we refer to the phrase as a concept, a framing, a framework, and a rhetorical device. Each use conveys a slightly different meaning and serves different purposes according to the needs of the discussion at hand. However, these notions are highly interrelated and reinforce the significance of the phrase "long tail" for LIS and the important need to deconstruct it through this literature review.

## The Methodology

This review initially emerged from a larger study about the data management practices and systems of astronomers (Stahlman, 2020). The original study explored the characteristics and prevalence of long tail data in astronomy, which is typically considered to be a "big data" field but is nevertheless impacted by the challenges of managing smaller and long tail data (Hanisch, et al., 2015). The literature review conducted for this earlier study is substantially extended here by taking a wider search of the literature about long tail data across disciplines, and by looking at how it emerged from and has been applied in the field of LIS. The review is divided into two main sections: a historical background that covers a wide variety of disciplines and an in-depth look into the LIS literature.

### Phase 1: Broad search

The first main section ("Broad Historical Overview") primarily reflects a traditional/narrative literature review approach (Paré & Kitsiou, 2017). We began this exploratory research with a Google Scholar search for the key phrase "long tail data", covering all possible disciplines, but limited to English-language journal articles and conference proceedings published between 2007 and 2022. With 1130 total search results, each result was screened for relevance and redundancy. We focused on papers that specifically reference long tail *scientific* datasets and collections, while excluding a vast body of literature about recommender systems, truth discovery, and modeling systems (where unlabeled or underrepresented data categories can introduce difficulties and bias).

  With a final initial bibliography of 153 papers, we further manually reviewed all papers, and selected 38 sources to be included in this review. We used the following inclusion and exclusion criteria in our decision-making to capture the most impactful and appropriate sources:

    *Redundance:* In many cases, multiple or similar versions of the same paper appeared in the bibliography (such as papers published as proceedings and subsequently journal article). In these cases, we selected the most relevant source for inclusion.

    *Relevance:* We focused on papers that address the central goals of this literature review to provide background information as well as information about long tail data case studies in specific disciplines. We favored papers that support the "narrative" purpose of this review to document the evolution of dialogue and practice over time.

Also, where it supported the discussion, we supplemented this bibliography with backward and forward snowballing (Wohlin, 2014) and our own expert knowledge of the literature to identify additional information while reviewing the papers and their references, to present a panoramic perspective on the history, issues, and discourse surrounding long tail data.

  A synthesis of the body of knowledge we gathered through this initial broad review is organized thematically into three broad categories identified through our qualitative assessment of the broader literature: 1) *Digital Libraries and Digital Curation*, which contextualizes the conceptual history of long tail data within the transition to digital information and collection-oriented thinking about data in the 1990's and 2000's; 2) *Cyberinfrastructure and Open Science*, which explores how these prior developments led to new computational resources and paradigms for data management and sharing; and 3) *The "Long Tail" of Scientific Data Across Contexts*, which looks at how different disciplines approach and discuss long tail data.

### Phase 2: Deep dive

  Following analysis of our initial search results, we discovered limitations, as some papers that we expected to be retrieved were not included in the results. Gusenbauer & Haddaway (2020) show that despite broad coverage, Google Scholar does not deliver replicable results and should not be used as principal search systems in a systematic review. While it is useful for exploratory research such as our Phase 1 broad search, lower precision reduces the reliability of Google Scholar for literature reviews. In order to compensate for this limitation, we devised a complementary search strategy that allowed us to more deeply analyze the literature, described as follows.

　　We discussed and compiled a list of papers that we consider to be seminal work about long tail data, assessed their contributions and impact, and looked at references to the papers over time to explore how the concept of long tail data percolated through LIS literature. We believe this methodological approach - broad searching complemented by a deep dive - has been very useful to take advantage of the coverage of Google Scholar while closely analyzing the literature within a domain we are familiar with. We believe this combinatory methodological approach could be implemented for other reviews with similar limitations as well.

　　The second main section ("An In-Depth Look into the LIS Literature") therefore presents a meta-analysis and offers an in-depth exploration of a concise sample of representative papers in LIS. Through our initial snowballing review and historical analysis, we identified seven highly referenced papers that we consider to be seminal for this topic in LIS, representing shifts in thinking about data management in the first decades of the 2000s, with implications for long tail data and small science. Using Google Scholar, we collected papers that cite those seven seminal papers and, after checking for relevance and redundancy, created a subset of the most popular (highly cited) papers aiming at including at least one paper per each seminal paper per year published between 2007 and 2022 (n=92). This subset of references to the seminal papers was analyzed qualitatively for deeper insight into the evolution of our topic of interest over time. Each paper in our sample was openly coded with descriptive phrases, and these phrases were grouped into thirteen overarching categories. We present and discuss the prevalence of these categories across our sample of literature over three time periods of interest: 2007-2011, 2012-2016, and 2017-2022.

### Broad Historical Overview

　　This section provides an overview of the history of scientific data management and curation, with a focus on the concept of long tail data and its impact on open science and reproducibility. Scientific or research data curation involves managing, preserving, and providing access to data for present and future use (Ball, 2010). Over the years, discussions about long tail data explored the best ways of curating such data, including convincing others to share data and integrating small data into larger collections to increase the benefit to scientific communities (Captur, et al., 2016; Vanderbilt & Gries, 2021; Wallis, et al., 2013). The dynamic and complex nature of data ecosystems, driven by the integration of both large and small data from multiple sources, reflects ongoing persistent challenges and themes in the field (Edwards et al., 2011; Ferguson et al., 2014).

### Digital Libraries and Digital Curation

Digital libraries and digital curation initiatives and practices have played significant roles in the management and preservation of long tail scientific data over the past decades. As Arms (2012) highlights, the "digital revolution" of the 1990s was a formative period for computing in libraries, setting the stage for modern data management and accessibility. At the same time, the evolution of digital libraries can also be traced back to much earlier visions for mechanized storing and retrieving of information (Bush, 1945; Lesk, 2012). In their goals of taking care of a wide variety of information assets, libraries have embraced technologies to meet changing information needs and to maintain relevance in the age of digital information (Borgman, 2000).

Digital *curation*[1] emerged from the digital library community in the early 2000's as a school of thinking, profession, and growing subdiscipline of LIS focused on long-term management of digital information assets, including a significant emphasis on long tail data (Beagrie, 2006; Higgins, 2011; Kouper, 2016). Albeit related to more general framings such as digital preservation and digital archiving, digital curation is about "maintaining, preserving and adding value to digital *research data* throughout its lifecycle"[1], with scientific communities playing an important role in developing current understandings and practices (Oliver & Harvey, 2016). Adding value is critical for long tail data as it involves not only preserving but also enhancing data that may not initially appear significant but can be vital for niche research areas and individual efforts. Importantly, digital curation consistently adopted and advocated for a lifecycle perspective on data management (Pennock, 2007; Higgins, 2008), with numerous data lifecycle conceptual models created to guide practices and systems for curation across research contexts and types and sizes of data collections (Weber and Kranzlmüller, 2019; Huang, et al., 2020; Stahlman, 2022).

Digital curation in the context of data and libraries grew out of concerns with publishing and preserving long tail data (Gold, 2010; McLure et al., 2014). By the early 2010s, digital data curation and the roles libraries play in it became a stable topic of research and practice (Hank & Davidson, 2009; Tibbo & Lee, 2012). At first, a relatively small number of research libraries had units and staff dedicated to working in digital data curation (Cox & Pinfield, 2014). Moreover, the efforts of library and scientific communities in data sharing and curation were initially rather separate, with venues where these efforts could be jointly discussed and synergized slowly emerging (Downs et al., 2015; Gray et al., 2002). Over time, the capacity of libraries to provide data curation services grew with many more libraries in the US, EU, and around the world developing capacity for supporting their researchers and providing data curation and management services (Kaushik, 2017; Yoon & Donaldson, 2019).

---

[1] https://www.dcc.ac.uk/about/digital-curation

Long-term curation of scientific data is as much a social challenge as a technical one, relying upon frameworks of policies, standards, roles, and responsibilities that exist to satisfy the diverse needs of stakeholders as well as the sustainability of digital data collections (Cragin & Shankar, 2006). Around the second decade of the 2000's, many funding agencies implemented data management planning requirements forcing researchers to think more carefully about data publication strategies and lifecycles (Smale, et al., 2020). At the same time, conversations surrounding curation of scientific data shifted towards addressing the sociotechnical challenges of incentivizing and supporting researchers in sharing their data and reusing the data of others (Borgman, 2012; Borgman, et al., 2016).

The issues of incentives and support were especially important for "small science" and long tail data which face clear drawbacks due to limited time and resources and risks of misuse or "scooping", but at the same time may benefit from opportunities for collaboration and enhanced visibility of scholarship through data sharing (Borgman, 2015; Wallis, et al., 2013; Wallis, 2014). Amidst rapid transitions to new hardware and software platforms throughout the 1990's and 2000's, attention also turned towards data "rescue" efforts to migrate digital data from legacy media formats (Wyborn, et al., 2015). Funding and lack of other resources and expertise were often among the challenges that affected the efforts of managing and preserving both newer and older data (Gonzalez & Peres-Neto, 2015; Griffin, 2017; Jahnke & Asher, 2012).

To illustrate the unequal distribution of financial and human resources in managing long-tail data, Palmer et al. (2007) and Heidorn (2008) drew inspiration from the Pareto principle (Juran, 1954) – the "law of the vital few", or the "80/20 rule" that described how about 80% of the land in Italy in the 19[th] century was owned by about 20% of the population. This principle was first adapted for LIS research by Trueswell (1969) to show that approximately 80% of library transactions represent approximately 20% of all items. This rule and the notion of a "long tail" have been used in various framings of data, including size (Ferguson, et al., 2014), scale (Borgman, et al., 2015), visibility and long-term sustainability (Liang, et al., 2010), findability (Curry & Moosdorf, 2019), and competitive value (Malik & Foster, 2012). Despite the variety of interpretations, the "80/20 rule" and "long tail" consistently serve as metaphors to highlight discrepancies between outcomes, such as data size and importance, and their causes, such as funding, curation, or quality.

　　　　Besides increased funding, standardization in data management has been seen as a way to promote openness, sharing, quality, and sustainability in long tail data and smaller communities, generating many clever acronyms such as TOP, FAIR, CARE, and TRUST (*CARE Principles for Indigenous Data Governance*, 2019; Lin et al., 2020; Nosek et al., 2015; Wilkinson et al., 2016). The widespread prevalence of these principles in conversations surrounding data represent a cultural shift towards practicing open science and developing shared understandings and norms surrounding data production, management, sharing, use, and reuse (Oliver et al., 2023a; 2023b). Similar to "knowledge infrastructures" (Borgman, et al., 2015; 2019) and "information ecologies" (Oday & Nardi, 2003; Baker & Bowker, 2007), these conversations encourage a systemic, networked understanding of the relationships between humans and artifacts and how they enable and advance scientific research and knowledge. Overall, the study and practice of long tail data curation are increasingly embedded within analogies such as culture, infrastructure, and ecology, demonstrating the persistent complexities of managing these data.

　　　　To summarize, digital libraries and digital curation practices have played a significant role in advocating for and managing and preserving long tail scientific data over recent decades. Starting in the 1990s, the emergence of digital libraries and digital curation specifically focused on small science and long tail data. Furthermore, the scholarship on digital and data curation have addressed both the social and technical challenges of data management, including the challenges of funding, human capacity, sustainability, and infrastructure for data sharing and reuse. The evolution of the latter, the challenge of creating and sustaining infrastructure for data management and open science, will be discussed in the next section.

### *Cyberinfrastructure and Open Science*

　　　　The so-called Fourth Paradigm of data intensive scientific discovery has marked scientific progress in the 2000's, further elevating the role of data in science and making a case for new tools for data capture, curation, and analysis (Hey, Tansley & Tolle, 2009). Often seen as the new paradigm for the age of big data, the fourth paradigm recognizes the challenges of long tail data as well. As Jim Gray pointed out (ibid.),

"When the data finally shows up in your computer, what do you do with all this information that is now in your digital shoebox? People are continually seeking me out and saying, 'Help! I've got all this data. What am I supposed to do with it? My Excel spreadsheets are getting out of hand!' So what comes next? What happens when you have 10,000 Excel spreadsheets, each with 50 workbooks in them? Okay, so I have been systematically naming them, but now what do I do?"

"E-science" was a term that appeared along with the term "cyberinfrastructure" to refer to a) ways of doing science that incorporated networking and computing and b) emphasize the role of technology in managing all the data that comes from a variety of sources (Alvaro et al., 2011). Amidst increasing recognition of the value of well-curated data, Heidorn (2011) suggested e-Science as an opportunity for libraries to play key roles in helping research communities manage their data, with a particular emphasis on smaller (i.e., "long tail") datasets that could take on new life through reuse.

In the early 2000's, the influential "Atkins Report" (Atkins, et al., 2003; Atkins, Hey & Hedstrom, 2011) responded to concerns about the growing volume and complexity of digital information and increasing investments in information technology (IT) by laying out a vision and set of recommendations for the U.S. National Science Foundation (NSF) to support development of overlapping cyberinfrastructures that would include archives for computational research and data. "Cyberinfrastructure" (CI) refers to the interconnected digital systems, networks, and resources that support advanced research and data-intensive scientific investigations (Stewart et al., 2010). It encompasses high-performance computing, data storage and management, software tools, visualization capabilities, and high-speed networks, playing a crucial role in enabling researchers to collect, analyze, share, and preserve datasets.

In response to digital data and computing needs, library communities in the 2000s engaged in their own e-science efforts that resulted in creating institutional repositories (IRs) to support digital scholarship and data archiving (Lynch, 2003). As the volume of digital research outputs increased, institutions recognized the importance of establishing repositories to preserve and disseminate scholarly works, and IRs were created as centralized platforms to store and provide access to research outputs, including publications, datasets, and other scholarly materials. Over the past decades, IRs and community data repositories have played an important role in addressing the challenges associated with long tail data (Cragin, et al., 2010; Reilly, 2012). Researchers, particularly those working in niche disciplines that do not have established scientific repositories, can contribute their datasets to IRs, ensuring their long-term preservation and facilitating discovery by other scholars (Keil, 2014).

Both generalist (institutional) and specialist (scientific) data repositories have evolved over time and expanded their services in supporting long tail data, although they grew asymmetrically and sometimes struggled to respond to researchers' needs (Murillo, 2020; Rodrigues & Rodrigues, 2012). While certainly maturing as a form of data publishing, repositories sometimes rely on the early practices of publishing scientific literature and do not respond quickly to the changing needs of scientific data sharing and preservation that require attention to versioning, software curation, the increasing multimodality of data, and other issues (Assante et al., 2016; Hedstrom et al., 2013).

To address these challenges, efforts to integrate CI and data repositories emerged as part of management, curation, and sharing of long tail data (Choudhury & Kunze, 2009; Mokrane & Parsons, 2014; Parsons et al., 2011). By leveraging CI resources, such as big data storage, networking, and compute, repositories can handle larger and more diverse datasets, facilitate metadata standards and search functionalities, and support both data and software preservation. Such synergies can potentially enable researchers to access and synthesize a broader range of data, collaborate across disciplines, and conduct innovative research by leveraging the long tail of diverse and specialized datasets (Dietze, et al., 2013).

Initiatives that merge CI and repository approaches often emphasize going beyond one institution and enabling support for a larger scientific community or interdisciplinary efforts. Thus, one of the earlier CI / repository efforts funded by the NSF DataNet program "The Sustainable Environment through Actionable Data (SEAD)" project was designed to support sustainability science by leveraging a social networking curation model and creating repository spaces that could be used both as working and sharing spaces to entice researchers to share their long tail data across a federation of IRs (Plale, et al., 2013; Akmon, et al., 2017). Another project within this program, DataONE, created a network of interoperable scientific data repositories that, in addition to sharing, aimed to provide a variety of services, including metadata, computation, and visualization services (Michener et al., 2011). Larger scale working groups and consortia have also been formed to improve collaborations and long tail data sharing across various disciplines, including geosciences, neuroscience, and astronomy (Lehnert, et al., 2012; Ferrini, et al., 2013; Hanisch, et al., 2015; Hsu, et al., 2015; Heidorn, Stahlman & Steffen, 2018; Thompson, et al., 2020).

E-science, CI, and institutional and scientific data repositories have been funded, developed, and expanded with goals of enabling open science, a term that signifies various ways of broadening access to knowledge and research outputs as well as the culture of openness and transparency with support of technology (Ramachandran et al., 2021). The transition towards open science has been marked by a change in how we perceive data, viewing data as assets with intrinsic value, rather than necessary ingredients or byproducts of research activities (NAS, 2018; 2019). Supporting long tail data in the context of open science often implies recognizing the diversity of data that is being created and developing flexible plans that allow both researchers and data curators to adapt to the available funding mechanisms and digital environments (Horstmann et al., 2017).

This section has highlighted the evolution of data management through the development of cyberinfrastructure that can offer solutions in managing the growing volume and complexity of digital information. These developments aimed to increase access, preservation, and reusability of data through enhanced integration, software preservation, metadata standards, and search functionalities, thereby enabling more effective data sharing and advancement of disciplinary knowledge. In what follows, we will discuss the developments of data management in specific scientific contexts.

### The "Long Tail" of Scientific Data Across Contexts

Heidorn (2008) described much long tail data as "boutique", where there may only be a few researchers interested in specific datasets but taken together these data are numerous. Callaghan (2020) similarly contrasted long tail "artisanal" with large-scale "industrial" big data production, where artisanal data are created by individual researchers or small teams and organized according to the needs of the data creators without consideration for extending the life of the data beyond their intended purpose. Across the sciences, some disciplines are more prone to generating such "long tail", "boutique", or "artisanal" data by virtue of their methods, instruments, and research practices. The geosciences, for example, have been leading discussions and action around managing long tail data, primarily because of the nature of the data they generate to study and understand Earth systems on long timescales through unique and local observations that can be later integrated across disciplines (Sinha, et al., 2013; Ramdeen & Poole, 2017; Stephenson, 2019; Stephenson, et al., 2020; Chen & Chen, 2020; Lenhert, et al., 2019).

Scientific communities such as geosciences often have their own initiatives to standardize data formats and metadata descriptions to improve long tail data discovery, interoperability, and usability (Cutcher-Gershenfeld, et al., 2016; Stamps, et al., 2020). Geoscientists have also been instrumental in the development of digital repositories specifically designed to handle long tail data, such as the Geological Society of America's (GSA) Data Repository, which was established as a service for journal article authors in 1974 and migrated online in the early 2000s and, in 2020 to Figshare (*Data Repository: GSA Publications Supplemental Materials*, n.d.), or the PANGAEA Data Publisher for Earth & Environmental Science (Schindler, et al., 2012). These activities not only benefit the geosciences but provide models and best practices for other disciplines grappling with similar data issues.

Ecological research communities similarly produce substantial amounts of long tail data through research practices that often involve individuals or small teams collecting field-based observations on specific ecosystems or species, often with specialized instrumentation and sensors that create unique challenges for data management (Borgman, et al., 2016; Easterday, et al., 2018; Bond-Lamberty, et al., 2021). These communities have also been influential in discussions about management of long tail data, as illuminated by Brooks, et al.'s (2016) discourse analysis of a scientific workshop, which identified structural challenges within the sciences, such as academic collaboration norms and financial incentives for long tail data management as impediments to data democratization efforts.

A recent special issue of *Ecological Informatics* highlighted progress that has been made on long tail data synthesis efforts over the years (Vanderbilt & Gries, 2021), while emphasizing a continued need for domain specific terminologies and common data formats to make data discoverable (Bond-Lamberty, et al., 2021), common repository tools to support workflows (Lenters, 2021; Huber, et al., 2021), and clear guidance for data citation (Agarwal, et al., 2021). Nevertheless, data repositories such as Dryad (White, et al., 2008; Greenberg, et al., 2009; Vision, 2010; He & Nahar, 2016), DataONE (Michener, et al., 2011; 2012; Allard, 2012), and the Knowledge Network for Biocomplexity (KNB) (Berkley, et al., 2009) emerged from ecological communities as common platforms for long tail data curation, preservation, and accessibility. Ecologists have also participated in efforts to standardize data formats and metadata descriptions, developing data standards including the Ecological Metadata Language (EML) to facilitate data sharing (Fegraus, et al., 2005).

Long tail data is also abundant in biological and medical research fields, comprised of unique datasets such as lab-based studies, patient records, genomic datasets, and even null findings (Captur, et al., 2016). Much clinical research data is small in scope with limited reproducibility and unpublished results that contribute to repetition of studies (Hanson, et al., 2020; Almeida, et al., 2021). Examples of successful sharing of raw data beyond the lifecycles of the original studies and published results include long tail data of resting state electroencephalogram (EEG) measurements in Cuba, which was converted into a large dataset for widespread use (Bosch-Bayard, et al., 2020). In the field of neuroscience, Ferguson, et al. (2014) emphasize the importance of mining and aggregating diverse data, while Hawkins, et al. (2020) note the importance of data publication tools and machine learning techniques to manage data variety. Similarly, Wyllie & Davies (2015) recommend data warehousing as a solution for reporting integrating sensitive patient-level data for healthcare epidemiology across organizations and regions. Examples of data repositories that support long tail data sharing in biological and medical research areas include GenBank for genetic sequences (Costa, et al., 2016), the Protein Data Bank for 3D structures of large biological molecules (Dutta, et al., 2009), and clinicaltrials.gov for information about clinical studies (Kirillova, 2012).

Other research areas represented in the literature about long tail data include materials science (Akmon, et al., 2011), hydrology (Yu, et al., 2020), scientific ocean drilling (Collier, et al., 2015), statistical research (Bahls & Tochtermann, 2013), and astronomy (Stahlman, Heidorn & Steffen, 2018; Stahlman & Heidorn, 2020). By contrast, humanities and social sciences disciplines are not represented in published literature referring to long tail data. However, these disciplines are nevertheless conducive to long tail data dynamics such as heterogeneous data formats and data sharing challenges, and especially considering the rapid growth of digital humanities research (Wang, 2018), may soon appear in the scholarly discourse alongside traditional sensor- and sample-based long tail data discussions. International contexts are also gaining relevance for discussions about long tail data, particularly as developing countries struggle to participate in research data management (Van Deventer, 2015; Patterton, et al., 2018; Stahlman, 2023).

This section has demonstrated that while sometimes described as "boutique" or "artisanal," long tail data collectively forms a substantial body of scientific data through efforts to manage and share it. Geosciences and ecological research have been pioneers in addressing the challenges of managing such data through the development of specialized digital repositories and metadata standards. Biological and medical fields also have dealt significantly with long tail data, developing specialized repositories and data management tools to facilitate data sharing and reuse. The discussion underscores the broader application of long tail data management across various scientific disciplines, emphasizing the growing importance of standardized practices and international cooperation, particularly in developing countries, to enhance global research data management.

### An In-Depth Look into the LIS Literature

To further understand the evolution of the approaches to long tail data, we performed a deeper analysis of the LIS literature. The main goal of this analysis was to examine how the discussions surrounding long tail data and approaches to it are evolving over time.

To perform the analysis, we identified a seed sample of seven publications that were foundational to these discussions in the 2000s and were among the first to argue for the role of information professionals (librarians and curators) in supporting the practices of scientific data sharing and preservation. The choice of these papers is inevitably subjective and many other papers from the LIS literature could have been added to the list. Our choice was guided by a combination of factors that included high citation ranks, years of publication, contributions to the LIS literature, and a range of topics addressed by the authors. The following papers were included in the list of seven (listed chronologically):

1. Cragin & Shankar (2006) explores scientific data collections in the context of distributed scientific work.
2. Borgman, et al. (2007) discusses the challenges faced by habitat ecology researchers in managing and organizing large amounts of data from embedded sensor networks, highlighting insights into data practices, collaboration between scientists and computer scientists, and the need for data policy and digital library architecture.

3. Palmer, et al. (2007) proposes a research agenda and study focused on understanding data practices and needs among environmental scientists to explore the potential for inter-institutional data coordination.

4. Heidorn (2008) introduces the notion of "dark data" discussed above.

5. Cragin, et al. (2010) describes perspectives of scientists on data sharing, emphasizing the need for data curation services to accommodate various data characteristics and sharing practices, with institutional repositories playing a crucial role in managing and mobilizing scientific research data.

6. Akmon, et al. (2011) highlights the need for archivists to recognize the value of science data and their role in data curation and collaborate with scientists in data curation.

7. Heidorn (2011) describes the role of libraries in managing and preserving digital data.

From these papers, we then searched for and saved all papers that cited them. For qualitative analysis we selected papers with the highest citation rank per year among references to each of the seed papers. The number of citations was recorded via Google Scholar. The paper referencing each seed paper with the highest number of citations was selected for qualitative review. Theses and books were excluded from qualitative analysis. After additional review for relevance and errors, the resulting dataset consisted of 92 papers published between 2007 and 2022 with a relatively even number of publications per year (see Figure 1 below).

**Figure 1**

*Publications that cited seed publications by year (N=92).*

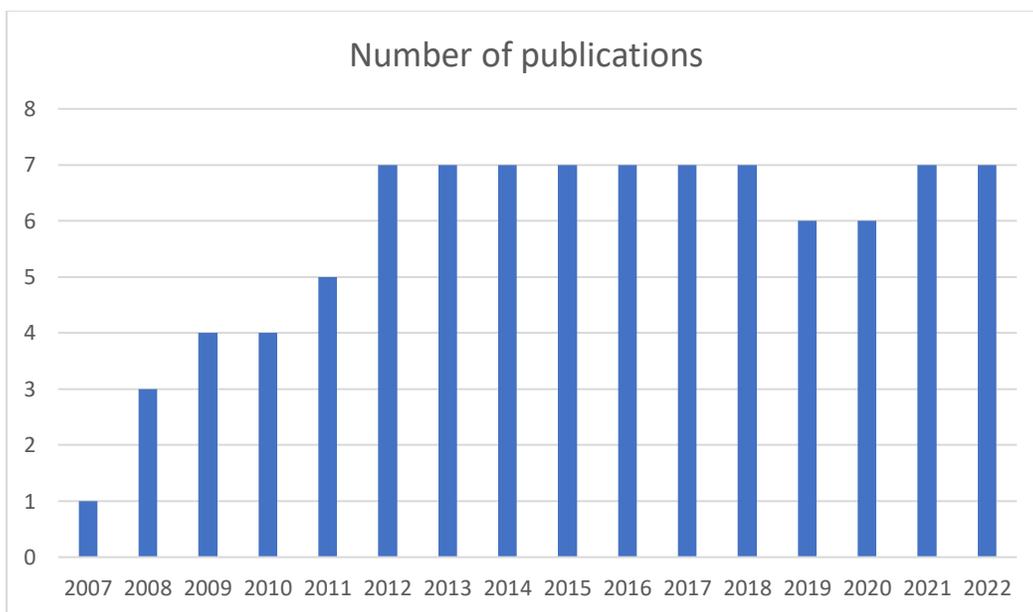

Four publications contained duplicates because they cited more than one seed publication. They were not excluded from the dataset because they contained context from each seed publication that was important for understanding thematic changes over time. For each publication we recorded its type, publication year, authors, title, publication venue, number of citations, disciplinary focus, themes, methods of research, and main findings or arguments.

Most of the publications in this sample were journal articles (82), with several conference papers (10). The publications came from a wide range of venues, including disciplinary venues (e.g., BioScience, Frontiers in Ecology and the Environment, and Nature Neuroscience), LIS venues (e.g., College & Research Libraries, International Journal of Digital Curation, and Library & Information Science Research), and broad focus venues (e.g., PLOS ONE, Scientific Data, and Industry and Innovation).

To better understand the recurring and emerging themes in the context of long tail data, each publication was coded with key phrases that described the main topics that the publication has addressed. The first round of coding involved free-text coding and resulted in a wide range of phrases related to the topic of long tail data, including data sharing, management, and reuse as well as digital libraries, open science, data quality, and so on. The total number of phrases was 257. These key phrases were further standardized into thirteen larger categories. Each of these categories speaks to various challenges and considerations prevalent in the discourse about long tail data. The distribution of themes was grouped by three temporal periods, 2007-2011, 2012-2016, and 2017-2022 (see Table 1 below).

**Table 1**

*Larger themes in publications by time periods (themes are arranged alphabetically, arrows indicate clear upward or downward trends).*

| Theme | Number and % of all themes | | |
|---|---|---|---|
| | **2007-2011** | **2012-2016** | **2017-2022** |
| Access and sharing ↗ | 5 (11%) | 14 (14%) | 16 (14%) |
| Analysis and visualization | - | 4(4%) | 4(4%) |
| Cyberinfrastructure | 3 (7%) | 8 (8%) | 3 (3%) |
| Education | 1 (2%) | 4 (4%) | 3 (3%) |
| Libraries ↗ | 1 (2%) | 2 (2%) | 5 (5%) |
| Management and curation | 6 (13%) | 17 (17%) | 13 (12%) |
| Metadata | 4 (9%) | 3 (3%) | 8 (7%) |
| Openness | 5 (11%) | 4 (4%) | 13 (12%) |
| Quality of data and research ↗ | 1 (2%) | 3 (3%) | 11 (10%) |
| Repositories and archives ↗ | 2 (4%) | 5 (5%) | 9 (8%) |
| Reuse ↗ | 1 (2%) | 3 (3%) | 9 (8%) |
| Scales and levels (e.g., big, small, dark) | 7 (16%) | 21 (21%) | 10 (9%) |
| Social and professional practices (e.g., collaboration, publishing, citation, communication) ↘ | 9 (20%) | 13 (13%) | 7 (6%) |
| **N publications (unduplicated)** | **17** | **35** | **40** |
| **N themes** | **45 (100%)** | **101 (100%)** | **111 (100%)** |
| **Avg N$_{themes}$ per publication** | **2.65** | **2.89** | **2.78** |

The table illustrates the seemingly relative stability and persistence of themes in the LIS literature as most of them remain in the discussions across all three time periods. Some themes, such as data analysis and visualization, education, and libraries had a rather small representation in the sample, five or less percent of all themes in each time period. Several themes showed a clear increase, e.g., access and sharing, libraries, quality of data and research, repositories and archives, and reuse. The discussion of social and professional practices that included references to the practices of collaboration, communication, publishing, and citation decreased over time. Other themes, such as cyberinfrastructure, metadata, and openness fluctuated over time. Below we provide examples of how some of these themes were addressed.

  ***Access and sharing, reuse.*** The themes that showed an upward trend in the examined publications focused on the issues of data access and sharing and data reuse. Even though the themes were relatively easy to separate during our coding as they used distinct terminology, the same papers were often coded with both themes because the theme of access and sharing was often used as a justification for data reuse. The earlier publications focused on such justifications to establish a research agenda within LIS that promoted both the practice of data sharing and the need to study barriers to it. Later on, the publications continued to link reuse with sharing, while examining data practices of specific scientific communities, such as ecology or neuroscience, with a number of papers based on the qualitative studies of the NSF-funded projects. Some studies provided evidence from systematic literature reviews and surveys and continued to demonstrate that data sharing remains uneven across scientific disciplines, facing many obstacles despite the proclaimed benefit of reuse. A few latest studies published in 2022 referred to the Findable, Accessible, Interoperable, and Reusable (FAIR) framework as a way to increase reuse and yet they still reported reluctance to share data in some disciplines due to lack of incentives and user-friendly tools.

   *Scales and levels.* One theme that had a rather large representation and then decreased over time (16%, 21%, and 9% across three time periods correspondingly) included the varying discussions of scales and levels of data, domains, and communities. Predictably, many publications used the term "long-tail data", but many others also referred to big data and big science, small or little science as well as web content and crowdsourced data. Often, the opposing size scales were used in the publication to argue for the value of long-tail data and the need to pay attention to it. Thus, the publications called for a deeper understanding of how scientists manage both large and long-tail data and use data in the context of "big", i.e., data-intensive, science, and "little" science, i.e., science that generates data of smaller sizes. Terms such as "scientific data", "research data", and "disciplinary data" were also used to discuss data management practices as a whole. Science and data practices were examined in publications through the lens of individual labs, larger communities, whole disciplines, or multidisciplinary collaborations. Through their use of varying scales and levels of data and scientific communities the authors continued to emphasize the importance of small datasets in the data universe and advocated for the role of library and archival communities in managing this type of research data.

   *Management and curation.* A similar trend of growth and then decline was observed in the theme of management and curation (13%, 17%, and 12% correspondingly). Earlier publications argued for the need to study the varying practices of individual researchers in order to provide curation and publishing services for long-tail data and develop institutional repositories that could offer such services. In fact, the management and curation theme in the early 2010s was often co-occurring with the theme of repositories and archives. In the mid-2010s the discussions of management and curation focused on specific communities, such as biodiversity scientists or marine biologists, expanding the focus from end-of-lifecycle data publication in institutional repositories to the broader challenges of making scientific data available, including building disciplinary data registries and databases, recognizing interdisciplinary collaborations, and tracking changes and updates to published datasets. As the theme becomes less frequent in the late 2010s and early 2020s, it nevertheless gains a global perspective through publications about research data management in non-US and non-EU contexts.

The themes of ***cyberinfrastructure*** and ***repositories and archives*** both addressed the issues of creating and maintaining computer, information, and communication technologies in support of data and knowledge production (Atkins et al., 2003). The former has fluctuated up and down over time (7%, 8%, and 3% correspondingly), while the latter has seen a steady increase (4%, 5%, and 8%). Similar to sharing and reuse these terms often co-occurred in our sample. One illuminating example demonstrates the terminological diversity when it comes to the discussions of how to conceptualize technological support for data sharing in the long tail (Marcial & Hemminger, 2010):

> Emerging SDRs [scientific data repositories] are building on recent ideas like collaboratories ..., shariums ..., and the cyberinfrastructure .. . They are utilizing practical tools from related domains, including institutional repositories (Eprints, DSpace, & Fedora), digital libraries and publishing worlds, and e-infrastructure (I Rule Oriented Data Systems or iRODS and gCube…

The authors of the publication above have used their own term "scientific data repositories" and referred to many other terms, some of which gained some prominence, which others, such as "sharium", have not. Despite these overlaps, cyberinfrastructure and repositories and archives were coded as separate themes because during our analysis it became clear that they were coming from two distinct scholarly contexts, one of them more oriented toward information science, while the other – toward library science.

The term "cyberinfrastructure" has often been used in the studies of specific scientific projects or domains and their data management practices and needs. It referred to broader digital technologies and tools to address collaborations over the research lifecycle, from data collection to analysis to sharing and reuse. It was also connected to "knowledge infrastructures" and "infrastructure studies", the areas that primarily draw on science and technology studies and conceptualize knowledge production and intellectual outputs as shared resources that are intended for public consumption (Frischmann, 2012).

The terms "repository" and "archives" had a narrower field of application. The term "repository" was primarily associated with efforts to build institutional repositories within universities to provide long-term archival and preservation of scholarly resources (Ware, 2004). In connection to data, the publications discussed various types of repositories, their possible classifications, and features of specific implementations (e.g., metadata, usage tracking, and sharing of computer code). The term "archives" was also connected to the efforts of long-term preservation of scholarly work. It was associated with a rather distinct movement within LIS research to identify and promote the role of archivists in long-tail data curation and preservation. Our seed paper by (Akmon et al., 2011) was among the prominent voices that argued for such a role. The subsequent citing publications, however, cited other seed papers as well while discussing approaches to archiving research data and storing it in institutional and disciplinary repositories. In the 2017-2022 period the publications expanded their discussion of repositories and archives to the understanding of researchers' perspectives on them and to the role of web resources in archiving data (both as a source of data and a storage functionality).

Finally, a small but growing theme in our literature sample addressed the issues of **quality** in data practices and scientific research. Within this growing theme the publications discussed how to improve research validity and integrity through such practices as replication, reproducibility, and interoperability. The authors described case studies from ecology, biology, environmental research, and hydrology where they have demonstrated how technological solutions, such as relational databases, online portals, metadata registries, or persistent identifiers helped to ensure research integrity and reproducibility. Human effort, particularly in data curation and management, has also been shown to affect data and research quality. At the same time, the authors found that preparing data for reproducibility requires additional resources and skills.

## Discussion and Conclusion

Through our historical review and qualitative analysis of influential papers, we have highlighted major themes and shifts in the broader discourse about "long tail data", a concept typically used to describe smaller scientific datasets and highlight the need for their curation through sustained effort and infrastructure. Theme evolution discussed above can be schematically presented in Figure 2:

**Figure 2**

*Evolution of long tail themes.*

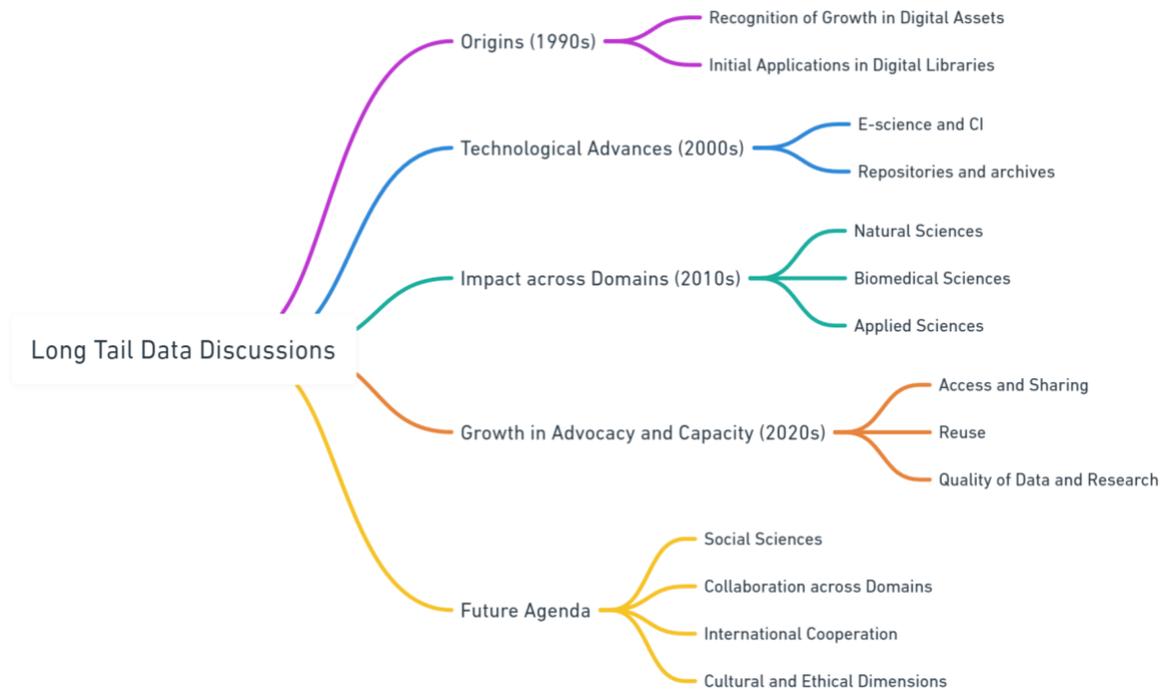

The term "long tail data" emerged in the LIS literature and grew through the efforts of scholars and practitioners who saw the need to support management and preservation of research data in its diversity of forms and sizes. Inspired by Anderson's (2007) long tail economics, which argued for the value of niche products and their contribution to the markets, Palmer, et al. (2007) and Heidorn (2008) introduced the concept of long tail to draw attention to challenges and opportunities in managing smaller data collections across the sciences.

Our analysis has shown that the long tail data concept remains relevant both as a rhetorical device and an ongoing way of conducting scientific research, even as science is rapidly evolving towards artificial intelligence applications and larger scale data archives and infrastructure. Much of the literature covered here discusses the unique data collection methods, instruments, and research practices that continue to be relevant for specific disciplines and areas of research, such as geosciences, ecology, neuroscience, or psychology. While efforts to collect and manage long tail data have been most successful as individual endeavors or small-scale collaborations, Brooks, et al. (2016) suggested constructing support and infrastructure around data collection types and methods rather than siloed disciplines, thereby increasing transdisciplinary and collaborative opportunities for long tail data.

The concept of "lifecycle", whether data lifecycle or research lifecycle, is prevalent in the literature and has been promoted to guide data management throughout research projects and beyond in accordance with typical archival principles such as preservation and appraisal (Pennock, 2007). The notion of a lifecycle speaks to fundamental challenges associated with managing long tail data over time. As shown throughout this review, long tail data are frequently generated for a project and not shared beyond the research team or laboratory, which puts an end to the lifecycle before the data can be used by others. However, given that long tail data are considered to be a consequence of inadequate funding and infrastructures, researchers often lack resources and support for sharing data – meaning that while the data may be dutifully published online, there may nevertheless be limited or poor-quality metadata and reduced capacity for discoverability and reuse.

Libraries and LIS communities have embraced the concepts of data sharing and open science and engaged in the studies of barriers to data sharing and the development of institutional repositories in support of small science and its data (Cragin et al., 2010). Meanwhile, research communities and agencies have become increasingly involved in data management collaborations around individual disciplines and research problems, leveraging public and private investments in computation and cyberinfrastructure resources for distributed data sharing and analysis (Atkins, Hey & Hedstrom, 2011). Overall, the concept of open science now often structures collective discourses about scientific research and the need to improve its processes and outcomes and regain public trust in science (Mirowski, 2018).

The topics of openness, access, and sharing have shown stable prominence in the discussions about long tail data, often fluctuating up and down in response to the data management requirements issued by the National Science Foundation and other agencies. These requirements were often used as a proverbial "stick" in attempts to rally support from institutions and individual scholars in developing data services and making curation and management a more prominent part of scientific processes. Despite these efforts and several notable successful examples, the barriers to both data sharing and wider data curation persist (Meyer, 2009; Tenopir et al., 2011; Houtkoop et al., 2018). Nevertheless, efforts to build curation and sharing services for long tail data continue to grow, often through efforts to develop diverse cyberinfrastructures and redefine support as distributed collaborative services with limited scope that help organize research data, find appropriate computational resources, promote data literacy, and train future generations of data scientists (Cox & Pinfield, 2014; Joo & Schmidt, 2021; Kim, 2021).

The thematic expansion in the long tail framing can be seen as both an indication of progress and a reflection of persistent challenges. One example that illustrates this progress / persistent challenges duality is the data ecosystem of the National Aeronautics and Space Administration (NASA), a United States government agency that supports a wide variety of facilities and research spanning aeronautics, earth and planetary science, astrophysics and space research and technology development. The scale and scope of NASA projects range widely between areas such as deep-sea exploration to space telescope construction and management. NASA's vast public data archives represent current best practices in openness, cyberinfrastructure, attention to data quality, and the development of services and platforms. At the same time, NASA research communities presently struggle with managing their data, particularly long tail data, in areas such as planetary science (Bristow, et al., 2021), astronomy (Heidorn, Stahlman & Steffen, 2018), and geosciences (Agarwal, et al., 2021). This implies that current approaches to data lifecycle management are not sufficient in supporting large heterogeneous communities of researchers and may need to be reconsidered, adjusted, or changed.

The long tail data approach is a relatively straightforward approach that did not yet make significant theoretical contributions to the LIS field, but nevertheless produced a large variety of work that described practices "on the ground". These include the individual practices of working with scientific data across a variety of disciplines and projects as well as the institutional practices of building cyberinfrastructure, promoting standards and guidelines, and adapting to the technological changes. The terminological diversity that we observed in the literature regarding how long tail data is described reflects the diversity of approaches and conceptualizations in what is being studied and how. For example, as highlighted above, researchers invoke a long tail distribution to describe challenges ranging from data sizes, which is a technical issue, to number of researchers working with datasets, which can be considered a social issue. Often, the notion of long tail is implied rather than explicitly stated.

Cultural shifts remain essential to produce sufficient incentives and recognition for both long tail and big data sharing, as well as fluidly integrating and reusing data across these contexts. Cultural factors, at many levels, govern the availability of data, what is done with data, and how data are organized and presented. Oliver, et al. (2022), define "data culture/s" as "the social, technical, and cultural characteristics, values and practices that influence/determine the nature of data production, generation, acquisition, cultivation, use, curation, preservation, sharing, and reuse by individuals, organizations, governments, and societies" (p. 1). All components of this framework as identified by the authors can be linked to the topic of long tail data. First, *cultures of data-related skills and attitudes* can influence and aid researchers in acquiring the skills and knowledge needed to integrate long tail data with open science ecosystems, where some degree of technical and organizational proficiency may be essential. Second, *cultures of data sharing*, which are often associated with specific academic disciplines, represent the attitudes and behaviors that foster data sharing amidst the significant societal influence of the Open Science and Open Government movements. Third, *cultures of data use/reuse*, also typically specific to scientific fields, are key to realizing visions for data sharing, which rely on reuse. Fourth, *data ethics and governance cultures* focus on leveraging perceived organizational value of data, as well as ethical decision making around data for the benefit of society, both of which are core arguments for long tail data management efforts.

A recent workshop "Changing the Culture on Data Management and Sharing" produced a series of recommendations that ring true for the wider scientific community, including cultivating skills and expertise across different types of institutions; incorporating data management and sharing as part of scientific workflows; encouraging funders to support implementation; encouraging researchers to see data management and sharing as fundamental to science and their reputations; and facilitating trust in data governance (NAS, 2021). Future research should target these areas as part of the larger agenda to examine scientific enterprises and their data practices.

### Looking ahead

The literature review has traced the evolution of the "long tail" concept for scientific data over several decades. Within the present context of the artificial intelligence revolution, data of all sorts are invaluable to train robust models. Data science and data management work are therefore increasingly important, and the potential for long tail data to contribute both granular and wide-reaching insights is also increasingly significant. While we have established that long tail data are persistently abundant, our review has also shown that challenges including metadata, infrastructure, incentives for sharing, and data integration have remained prevalent over time. This points to a pertinent need for consideration and focused discussion about the future of long tail data management. Amidst vast amounts of data generated across disciplines, long tail datasets have a key role to play in advancing scientific research by revealing patterns and insights that may be overlooked in larger data collections.

LIS archival perspectives remain relevant to the further evolution and future directions of discussions about long tail data. Digital curation emerged from two LIS subdisciplines - archives and records and digital preservation - as a distinct field of research and practice (Higgins, 2008), and over time, this field largely catalyzed discussions about long tail data. While discourse surrounding this concept has since percolated across scientific disciplines as we have illustrated, expertise in curation and preservation remains essential, and as intermediaries between researchers and information, libraries continue to have a key role to play as hubs for facilitating data access and advocating for long-term preservation.

Emerging perspectives in archival studies can also inform future discussions about long tail data, as the archival literature increasingly grapples with data-related challenges and opportunities, while archives of all sorts can be viewed as data (Mordell, 2019). For example, Currie & Paris (2018) show that data preservation can be a form of political activism, informing the larger context of archival activism research. Particularly as humanities and social science disciplines demonstrate long tail dynamics with smaller, dispersed data collections, power dynamics and political implications of decisions and methods for archiving community-oriented data are important areas for consideration. On the technical front, archival lenses have turned towards the ethical and sociocultural implications of data science methods (Feinberg, 2022) and datasets used to train AI (Desai, et al., 2024), which are essential considerations for the long tail concept as well. While archival research has long studied affect in terms of information interactions and experiences (Cifor, 2016), Stahlman (2022) recently pointed to an opportunity and need for further research on the role of affect in long tail data management. Finally, computational archival science is emerging as a framework for research and education that integrates traditional data curation and archival practices (Fenlon, et al., 2019).

Overall, innovative and holistic approaches are essential to bridge knowledge domains and work towards ongoing integration of long tail data into mainstream data infrastructures and practices across fields. As we have shown through our literature review, these unique and often-hidden datasets contain a wealth of information that can empower researchers to test theories, explore case studies, and enhance models. They can also lead to currently unanticipated discoveries in the future. Managing and integrating long tail data effectively should be an essential goal and a shared task for LIS fields and across disciplinary contexts to harness the full potential of scientific research over time. We hope that our literature review about the evolution of the "long tail" concept up until this moment in time can inform these future discussions and efforts.

### *Limitations*

This literature review has several limitations. First, it does not have comprehensive coverage of the literature. While such a limitation is inherent in any literature review, our use of Google Scholar in literature search and compilation could have resulted in a less systematic approach compared to the use of more common search approaches, such as the use of Web of Science. Use of Google Scholar proved to be more productive in the context of fuzzy and overlapping terminology present in the literature about long tail data. Nevertheless, lack of transparency in Google Scholar indexing and ranking techniques somewhat undermined the representation of publications in our results. While we addressed this by conducting a deep dive into seven seminal papers and their references, some relevant sources may have been overlooked through this combinatory methodological approach.

Second, for the Phase 2 deep dive, we deemed seven papers "seminal" and worthy of close inspection. This determination, while based on our expertise and experience in the field of data curation, is subjective. Third, by conducting our Phase 1 search for the phrase "long tail data", variations and alternate terms were omitted from the search. Although our Phase 2 deep dive did retrieve alternate related framings – such as "dark data", "small data", etc. –we were primarily interested in deconstructing the multifaceted descriptor "long tail" in terms of data. Finally, while we briefly discuss the international context for long tail data in the section "The 'Long Tail' of Scientific Data Across Contexts", the breadth of international trends was not fully explored, as we focused on English-language papers in both Phase 1 and Phase 2 searches, particularly as the field of data curation originated within and across international English-speaking networks. The international trends in long tail data management are a worthy topic for a future review or study.

## Acknowledgements

We are grateful to Miriam Kim for assistance with literature analysis.

---

[1] https://dcc.ac.uk/about/digital-curation